\documentclass{aa}  
\usepackage{graphicx}
\usepackage{amssymb}
\usepackage{txfonts}
\usepackage{mathrsfs}
\usepackage{array}
\usepackage{natbib}
\bibpunct{(}{)}{;}{a}{}{,}   % to follow the A&A style
\begin{document}
   \title{Reddening and metallicity maps of the Milky Way bulge from VVV and 2MASS\thanks{Based on observations taken within the ESO VISTA Public Survey VVV, Program ID 179.B-2002}}
   \subtitle{II. The complete high resolution extinction map and implications for Bulge studies}
   
   \author{O. A. Gonzalez$^{1}$  \and M. Rejkuba$^{1}$ \and M. Zoccali$^{2,3}$ \and E. Valenti$^{1}$ \and D. Minniti$^{2,4}$ \and M. Schultheis$^{5}$ \and R. Tobar$^{1}$ \and B. Chen$^{5}$}
   
   \offprints{O. A. Gonzalez}
   \institute{ $^{1}$European
   Southern Observatory, Karl-Schwarzschild-Strasse 2, D-85748 Garching,
Germany\\ \email{ogonzale@eso.org; mrejkuba@eso.org; evalenti@eso.org}\\
   $^{2}$Departamento    Astronom\'ia    y Astrof\'isica,
   Pontificia Universidad  Cat\'olica de Chile,  Av. Vicu\~na Mackenna
   4860,         Stgo.,         Chile\\         \email{mzoccali@astro.puc.cl;
dante@astro.puc.cl}\\
   $^{3}$INAF-Osservatorio Astronomico di Bologna, Via Ranzani, 1-40127 Bologna, Italy\\
   $^{4}$Vatican Observatory, V00120 Vatican City State, Italy\\
   $^{5}$Institut Utinam, CNRS UMR6213, OSU THETA, Universit\'e de Franche-Comt\'e, 41bis avenue de l'Observatoire, 25000 Besan\c{c}on, France\\
}
   \date{Received ; Accepted }

   \keywords{Galaxy: bulge - ISM: dust, extinction - stars: abundances}
  
\abstract  
%  context (optional)  
{The Milky Way bulge is the nearest galactic bulge and the best laboratory for studies of stellar populations in spheroids based on individual stellar abundances and kinematics. These studies are challenged by the strongly variable and often large extinction on a small spatial scale.}
%  aims
{We use the Vista Variables in the Via Lactea (VVV) ESO public survey data to measure extinction values in the complete area of the Galactic bulge covered by the survey at high resolution.}
% method
{We derive reddening values using the method described in Paper I. This is based on measuring the mean $(J-K_s)$ color of red clump giants in small subfields of $2' \times 2'$ to $6' \times 6'$ in the following bulge area: $-10.3^\circ \leq b \leq +5.1^\circ$ and $-10.0^\circ \leq l \leq +10.4^\circ$. To determine the reddening values $E(J-K_s)$ for each region we measure the RC color and compare it to the  $(J-K_s)$ color of RC stars measured in Baade's Window, for which we adopt $E(B-V)=0.55$. This allows us to construct a reddening map sensitive to small scale variations minimizing the problems arising from differential extinction.}
% results
{The significant reddening variations are clearly observed on spatial scales as small as $2'$.  We find a good agreement between our extinction measurements and Schlegel maps in the outer bulge, but, as already stated in the literature the Schlegel maps are not reliable for regions within $|b| \la 6^\circ$. In the inner regions we compare our results with maps derived from DENIS and Spitzer surveys. While we find good agreement with other studies in the corresponding overlapping regions, our extinction map has better quality due to both higher resolution and a more complete spatial coverage in the Bulge. We investigate the importance of differential reddening and demonstrate the need for high resolution extinction maps for detailed studies of Bulge stellar populations and structure.}
% conclusion
{We present the first extinction map covering uniformly $\sim 315$ sq.\ deg.\ of the Milky Way bulge at high spatial resolution. We show that, in a 30 arcmin window, at a latitude of $b=-4^\circ$, which corresponds to often studied low extinction window, the so called Baade's Window,  $A_{K_s}$ values can vary by up to 0.1 mag. Larger extinction variations are observed at lower Galactic latitudes. The extinction variations on scales of up to 2'--6' must be taken into account when analysing the stellar populations of the Bulge.}
             
\authorrunning{Gonzalez et al.}
\titlerunning{High resolution extinction map of the Milky Way bulge}

\maketitle

%
%________________________________________________________________

\section{Introduction}

The Bulge is one of the major components of the Galaxy and it can be studied at a unique level of detail, in comparison to those of external galaxies, thanks to its proximity which enables properties of individual stars to be measured. Bulge stars hold the imprints of how our Galaxy formed and evolved. The Bulge star formation and chemical enrichment history can be investigated in detail through analysis of stellar ages and chemical abundances \citep[][]{mcwilliam94,zoccali03,fulbright07,rich_origlia07,lecureur07,zoccali08,clarkson08,brown10,bensby10,bensby11,johnson11,gonzalez11}. 

Our current understanding of the Milky Way Bulge structure and stellar content
can be summarized as follows: i) the Bulge structure is dominated by the Bar \citep{stanek94} which appears to have a peanut or X-shape in the outer regions \citep{mcwilliam10,nataf10,saito11}; ii) there is a radial stellar metallicity gradient \citep{zoccali08, johnson11} that most likely flattens in the inner regions \citep{rich_origlia07}; iii) in addition to the population associated with the bar, there are hints for a second chemically and kinematically distinct component \citep{babusiaux10, hill11, gonzalez11,bensby11}; and iv) the bulge is dominated by old stars \citep{zoccali03,brown10,clarkson11}, although there are indications for a possible younger population of metal-rich stars as traced by Bulge microlensed dwarfs \citep{bensby11}. However, the general view of these properties, based on a more complete coverage, is expected to provide the final picture of the Milky Way bulge formation history. Several projects aiming to obtain this general view of the Bulge are either ongoing or planned for the near future \citep[e.g.][]{vvv10,feltzing11,ness12,apogee12}. The interpretation of these observations is challenged by the variations in the dust extinction properties on small scales towards the different Bulge regions. 

The currently available extinction maps of the Bulge regions have different resolutions and coverages. The most commonly used extinction maps were those of \citet{bh82}, later superseded by  Schlegel maps \citep{schlegel98}. The Schlegel extinction maps are full-sky maps of the dust color temperature based on IRAS and DIRBE experiments which are then normalized to E(B-V) values using a calibration of colors of background galaxies. Unfortunately, these maps suffer from large uncertainties in regions towards the Bulge. Specifically, as stated in the appendix C of \citet{schlegel98}, the temperature structure of the Galaxy towards latitudes $|b|<5^\circ$ is not well defined and contaminating sources have not been completely removed. Furthermore, as we show below, the resolution of these maps is too coarse. As a result, the Schlegel reddening values towards the Bulge are unreliable. 

Clearly, an extinction map covering $|b|<5^\circ$, with sufficient resolution to resolve the small spatial scale variations of extinction, is important for an understanding of the general properties of the inner Galactic regions. Using red giant branch (RGB) stars from the near-IR photometric survey DENIS, \citet{schultheis99} provided a high resolution (2') map in the inner regions of the Bulge ($|b|<2^\circ$). This map provides the sufficient resolution to analyse such inner Galactic regions, but its coverage is too limited to permit the study of the {\it global} Bulge stellar populations and structural properties. 

More recently, \citet{marshall06} provided a full 3D extinction map of the Bulge. 3D maps are indeed of high importance for Galactic studies towards the inner Galaxy as they hold the additional distance information. However, in order to build 3D extinction maps, a sufficient number of stars is required in each resolution element, which implies that only a modest resolution can be achieved. The method in \citet{marshall06} is based on a comparison of 2MASS photometry to the Besan\c{c}on model. At low Galactic latitudes ($|b|<3^\circ$) 2MASS suffers strongly from incompleteness and blending due to too low resolution and high stellar density, limiting the ability to derive correct star counts and extinction.  The resolution of Marshall's 3D extinction map is 15', which is again too coarse to fully overcome the differential reddening problems at lower latitudes $|b|<4^\circ$. 

Other important extinction maps are available with a partial coverage of the Bulge such as those of \citet{kunder08} and \citet{sumi04}. However, for detailed studies of the Bulge, a high resolution and homogeneous extinction map covering the complete area of interest is missing. The deep and near-IR photometry of the VVV ESO public survey of the Bulge \citep{vvv10} provides the ideal dataset to create such a map. In \citet[][Paper I]{gonzalez11b} we showed an effective technique based on the color of red clump (RC) stars to trace the effects of extinction. In this article, we extend our analysis to the complete Bulge region covered by the survey and derive the first complete extinction map of the Bulge, using a homogeneous population, namely the red clump, at a resolution of 2'--6' (Sect.~\ref{sect:map}). We compare our map to extinction maps in the literature in the corresponding overlap regions (Sect.~\ref{sect:comparison}), and we discuss the importance of high resolution extinction map in future Bulge studies (Sect.~\ref{sect:implications}). 

%__________________________________________________________________
\section{Determination of $A_{K_s}$ values}
\label{sect:method}

\subsection{VVV survey data}

In this work we use the $\sim 315$ sq.\ deg.\ coverage of the Galactic bulge from the VVV survey. This corresponds to the first data release (DR1) of the survey, as described in \citet{saito12}.  In Paper I, we presented a detailed description of a method used to produce the multiband ($J$, $H$, and $K_s$) catalogs required for our global study of the Bulge stellar populations. Thus, we now provide only a brief summary of the adopted procedure. Single band catalogs for each VVV image, so-called tiles, are produced at the Cambridge Astronomical Survey Unit (CASU). These catalogs are then individually calibrated to the 2MASS photometric system by comparison of high quality stellar detections from both catalogs, in the magnitude range between $12<K_s<13$ mag where both catalogues overlap. Sources brighter than $K_s=12$ mag in each VVV tile are removed and are then replaced with 2MASS stellar detections, that have been flagged with high quality photometry. This ensures the required photometric agreement between the 2MASS and VVV surveys and the correction of saturation at the bright end ($K_s<12$) of the VVV catalogs.  These single band catalogs are then matched using STILTS \citep{taylor06} to produce our final multiband catalogs. 
 
\subsection{Mean color of the Bulge red clump}

Studies of the chemical abundances and age measurements of the Bulge towards Baade's Window, have shown that the Bulge is predominantly old ($\sim$10 Gyr) \citep{ortolani+95,zoccali03,brown10,clarkson11} and that it shows a broad metallicity distribution with a peak near Solar metallicity \citep[][]{mcwilliam94,zoccali08, johnson11}. Stellar evolutionary models can thus provide specific constraints on the expected intrinsic color of stars along the different evolutionary stages. The observed color can then be compared to that expected from models and its difference can be linked to an external factor, such as extinction. The metallicity of the Bulge is not spatially uniform, but it shows a gradient along its minor axis \citep{zoccali08}. Furthermore, the extent of these gradients towards other regions is not yet clear. For this reason, uncertainties may arise from direct comparison of the observed red giant branch (RGB) stars' colors with respect to those expected from models. Fortunately, this effect can be minimized by the use of RC giants, as the mean color of these stars has a small dependance on these parameters.  
\begin{figure}
\begin{center}
\includegraphics[width=0.46\textwidth]{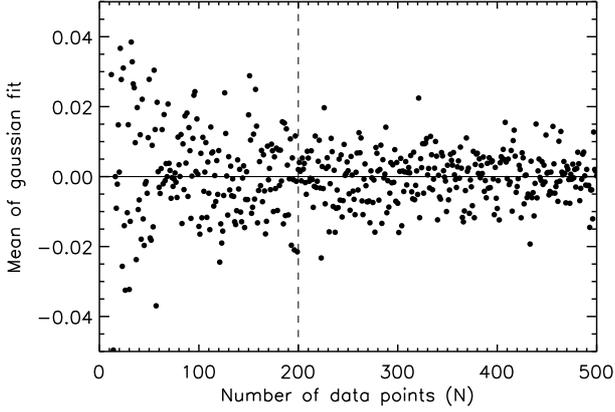}
\caption{The mean value obtained from a Gaussian fit to a set of N data points randomly distributed following a Gaussian distribution centered on zero (solid horizontal line) and with a sigma of 0.10. The dashed vertical line shows the limit of 200 points adopted as the minimum number of stars in each subfield.}
\label{nstats}
\end{center}
\end{figure}

\begin{figure}
\begin{center}
\includegraphics[width=0.46\textwidth]{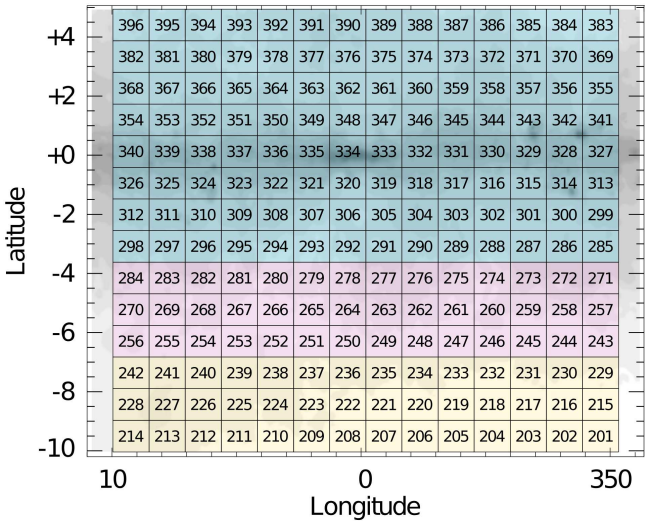}
\caption{VVV coverage of the Bulge in Galactic coordinates. Each small square shows the coverage by individual $1^\circ \times 1.5^\circ$ tile from the VVV. Tile numbering for the VVV survey is described in \citet{saito12}. The different color shading of the tiles depends on the spatial resolution (2', 4' and 6', respectively) used to derive reddening from the mean color of the red clump stars (see text for details).}
\label{resolution}
\end{center}
\end{figure}

In Paper 1 we used the multiband VVV catalogs to determine extinction values by measuring the mean ($J-K_s$) color of the RC stars and comparing it to that of a region of know extinction, named Baade's Window.  A selection box in the CMD of each tile is chosen to include only Bulge red clump stars. The limits of this box vary for each line of sight depending on the amount of extinction and were therefore inspected by eye and modified accordingly. A Gaussian fit to $(J-K_s)$ color distribution is then used to determine the mean color of the red clump. For a $4' \times 4'$ region in Baade's Window centered on coordinates $\alpha_{2000}=18:04:51.2$ and $\delta_{2000}=-30:03:26.5$, VVV photometry yields a $(J-K_s)=0.96$ mag. Using the measured extinction of this field of $E(B-V)=0.55 \pm 0.01$ mag \citep{sumi04,kunder08,zoccali08}, we obtain a dereddened mean color of $(J-K_s)_0=0.68$ for the RC stars in this region. We use this value as a reference for the intrinsic mean color of the RC, and its difference with respect to the measured RC color in any other region will be an indicator of extinction, meaning $E(J-K_s)$.

\section{The complete high resolution extinction map}
\label{sect:map}

\begin{figure*}
\begin{center}
\includegraphics[width=0.92\textwidth]{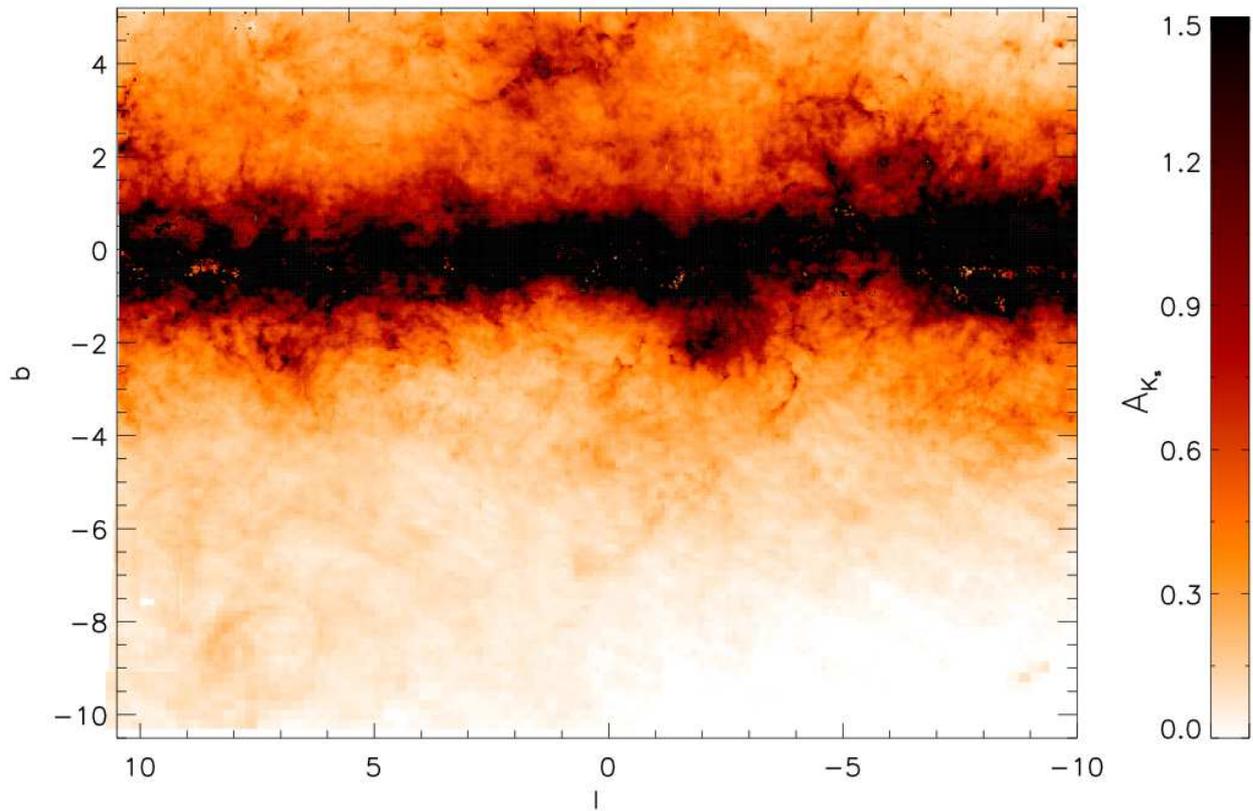}
\caption{Extinction map of the Galactic Bulge for the complete region covered by the VVV survey. The $A_{K_s}$ values are computed from  $E(J-K_s)$ measurements assuming \citet{cardelli89} extinction law for all tiles. At $A_{K_s}$ values larger than 1.5 mag the color scale saturates. The details of the extinction variation in the inner highly extinct regions, where $A_{K_s}$ reaches up to 3.5 mag, are better seen in the upper panel of Fig.~\ref{compmap}.}
\label{fullmap}
\end{center}
\end{figure*}

We computed the mean RC $(J-K_s)$ color in small subfields towards the complete Bulge region of VVV. We adapt the resolution to provide a sufficient number of RC stars ($>200$) to populate the color distribution with enough statistics, and at the same time to minimize the effect of differential extinction as discussed in Paper I. The minimum number of 200 stars for each subfield was determined following the result shown in Fig.~\ref{nstats}. The vertical axis of Fig.~\ref{nstats} shows the mean value obtained from a Gaussian fit to a set of N random values that follow a Gaussian distribution centered on zero and with a sigma of 0.10. The vertical dashed line shows the limit of 200 stars. The dispersion on the mean value from the Gaussian fits increases significantly for a number of data points below this limit. The resulting resolution of the complete map depends on latitude, as shown in Fig.~\ref{resolution}. For the latitude range $-3.5^\circ \la b \la +5^\circ$ the density of Bulge stars allowed the highest resolution of 2' to be used. As the density of Bulge stars drops at higher distances from the plane, the resolution had to be reduced to 4' for $-7^\circ \la b \la -3.5^\circ $ and to 6' for $-10^\circ \la b \la -7^\circ$.  The mean measured RC color is compared to the intrinsic value measured in Baade's Window to derive the reddening value $E(J-K_s)$. From $E(J-K_s)$, the $A_{K_s}$ extinction values are obtained by applying a specific extinction law. Figure~\ref{fullmap} shows the complete $A_{K_s}$ extinction map for the Bulge VVV region based on \citet{cardelli89} extinction law. The extinction variations in the inner $2^\circ$ from the Galactic plane are better appreciated in the upper panel of Fig.~\ref{compmap} where the highest values of extinction reach $A_{K_s}=3.5$.

In a small number of regions  the differential reddening is too strong even over 2' resulting in large errors for reddening measurements, or even in extreme cases preventing the identification of RC feature. Some of these regions are visible as small white dots/squares in the upper panel of Fig.~\ref{compmap}. They are less obvious in Fig.~\ref{fullmap}, because of the resampling in the inner most regions in the presentation of the figure. All these regions are clearly identified as having large errors in our extinction database. 

Our extinction maps show clearly the small scale variations, as produced by the strong dust features traced, in particular at low latitudes. These features are now seen in great detail thanks to the high resolution and large coverage of our study.

\subsection{A note about the extinction law}

At present, there is no real consensus on which is the correct extinction law to be used for studies towards the inner Galactic bulge. Reddening maps are related to a given extinction law which can vary between different studies.  Recent results suggest that extinction properties might vary in different locations and possibly depend on the amount of extinction and dust properties. In particular, \citet{nishiyama06,nishiyama08,nishiyama09} presented a detailed study of the interstellar extinction law towards the inner regions of the Bulge ($|l|<2^{\circ}$, $|b|<1.0^{\circ}$) concluding that this varies significantly depending on the line of sight.

\citet{nishiyama06} also provided ratios of total to selective extinction which differ from a $R_V\sim3.1$ standard extinction law as obtained towards other regions. 

The problem of the correct extinction law towards the Galactic bulge is not adressed here, but will be part of another article (Chen et al. 2012, in prep) based on GLIMPSE and VVV datasets. However, in order to obtain extinction values from the reddening map presented here, $E(J-K_s)$ values need to be transformed into an absolute extinction through a given extinction law. Here, we decided to adopt two different extinction laws. The first is the more commonly used extinction law from \citet{cardelli89}. Adopting 1.240, 1.664, 2.164 micron as the effective wavelengths of a K2 giant (typical of red clump stars) for 2MASS from \citet{indebetouw05}, the coefficients corresponding to the \citet{cardelli89} extinction law are:
\begin{equation}
A_J=1.692 E(J-K_s)
\end{equation}

\begin{equation}
A_H=1.054 E(J-K_s)
\end{equation}

\begin{equation}
A_{K_s}=0.689 E(J-K_s)
\end{equation}
The second is the extinction law from \citet{nishiyama09}:
\begin{equation}
A_J=1.526 E(J-K_s)
\end{equation}

\begin{equation}
A_H=0.855 E(J-K_s)
\end{equation}

\begin{equation}
A_{K_s}=0.528 E(J-K_s)
\end{equation}
which has been determined, for the 2MASS photometric system, in the inner regions of the Bulge closer to the Galactic plane.

\begin{figure}
\begin{center}
\includegraphics[width=0.48\textwidth]{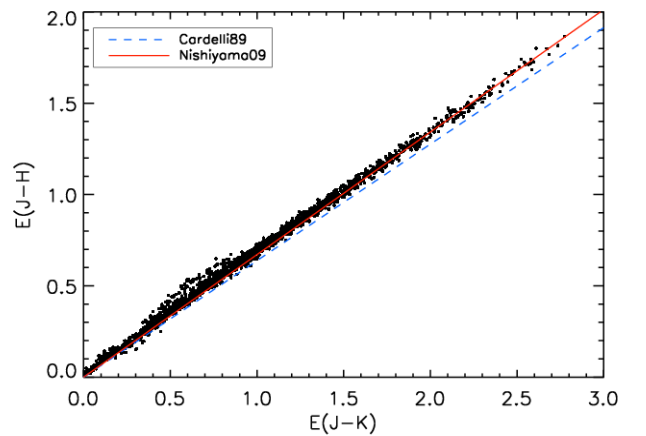}
\caption{$(J-K_s)$ and $(J-H)$ color difference between our control field and those measured in the subfields of tiles b317, b303 and b275. The size of the subfields corresponds to the same resolution described in Fig.~2. The blue dashed line shows the relation $E(J-H)=0.638E(J-K_s)$ corresponding to the extinction law from \citet{cardelli89} and red solid line to $E(J-H)=0.671E(J-K_s)$ from \citet{nishiyama09}}
\label{extlaw}
\end{center}
\end{figure}

\begin{figure}
\begin{center}
\includegraphics[width=0.43\textwidth]{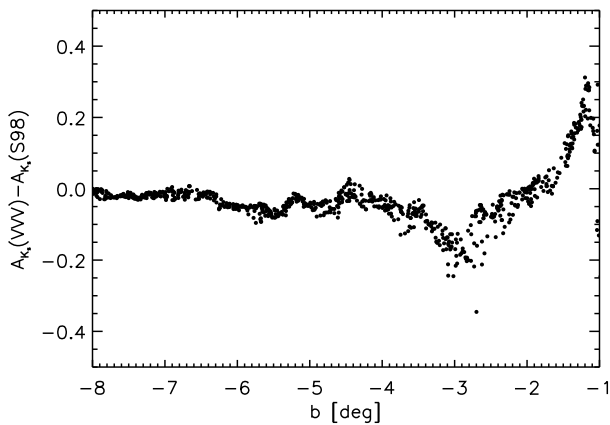}
\caption{Difference between the $A_{K_s}$ values obtained in this work and those of \citet{schlegel98}, as a function of Galactic latitude, for 1000 randomly distributed $30' \times 30'$ regions between ($-8^\circ <b<-1^\circ$).}
\label{schlegel}
\end{center}
\end{figure}

In order to test which extinction should be prefered, we performed the following exercise. We measured the colors $(J-K_s)$ and $(J-H)$ in our 4' control field in Baade's window, where we find mean values of $(J-K_s)=0.96$ and $(J-H)=0.73$. We then measure these colors for a set of regions with different latitudes and compare them to those in our control field. The $E(J-K_s)$ and $E(J-H)$ color difference for each line of sight are shown in Fig.~\ref{extlaw}. The relation between these color differences can be obtained from the corresponding coefficients of each extinction law. Our measurements are in good agreement to those with the \citet{nishiyama09} extinction law, where $E(J-H)=0.671E(J-K_s)$. However, since the implications discussed in Sect.~5 do not depend strongly on the adopted extinction law, we proceeded our analysis following the more commonly used law from \citet{cardelli89}.

\subsection{BEAM calculator}

As described in the following, our high resolution extinction map of the Bulge provides a valuable tool for future studies. In order to make it available to the community, we developed a web based tool named BEAM (Bulge Extinction And Metallicity) calculator\footnote{http://mill.astro.puc.cl/BEAM/calculator.php}. This tool, provides the user with an easy access to our measurements, both for the extinction values presented in this article, as well as for the photometric metallicities obtained in Paper I. The complete metallicity map and its analysis will be subject of the forthcoming Paper III (Gonzalez et al. 2012, in preparation). 

The tool allows the user to input a set of coordinates and the corresponding size of the field of interest, retrieving the mean extinction ($A_{K_s}$) towards these set of coordinates. The extinction values are computed using either \citet{cardelli89} or \citet{nishiyama09} extinction law. However, the $E(J-K_s)$ values are always provided as well, so that the user can adopt any other extinction law of preference.

\section{Comparison with existing extinction maps of the Bulge}
\label{sect:comparison}

\begin{figure*}
\begin{center}
\includegraphics[width=0.90\textwidth]{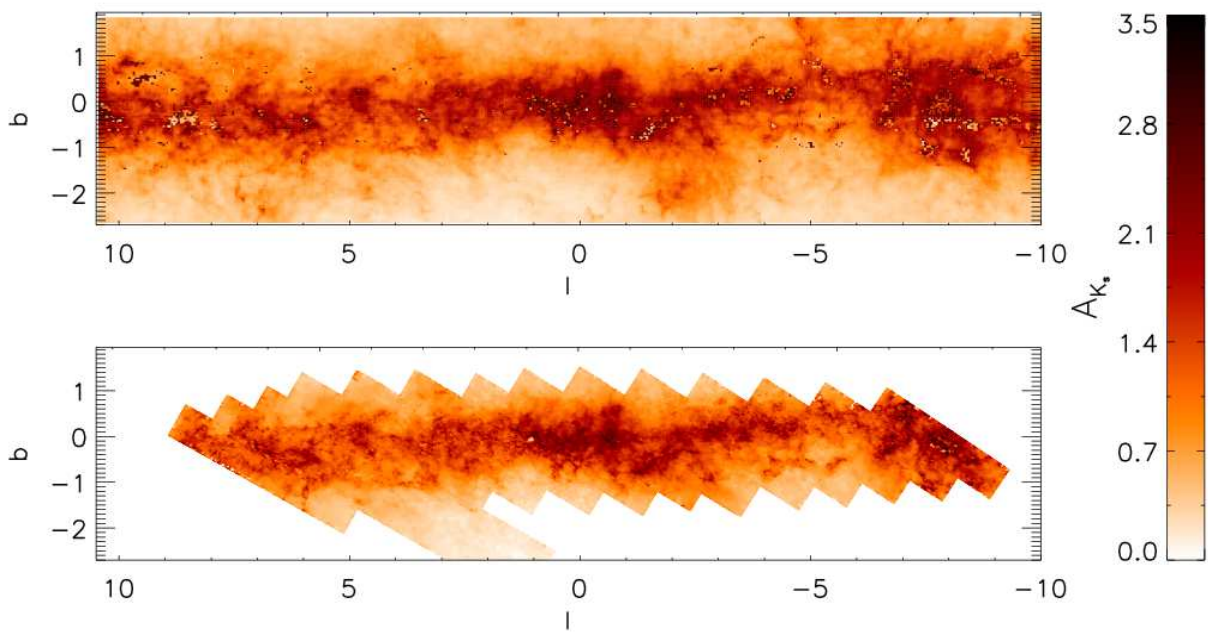}
\caption{The upper panel shows the inner $\sim 4^\circ$ region around the Galactic plane of our VVV extinction map. $A_{K_s}$ values are based on \citet{cardelli89} extinction law. Part of this region was also covered by the DENIS survey, used in \citet{schultheis99} to build an extinction map shown in the lower panel.}
\label{compmap}
\end{center}
\end{figure*}

Other studies have addressed the problem of extinction towards the Bulge. Reddening maps providing a similar Bulge coverage as the one presented here are the 3D maps of \citet{marshall06} and the \citet{schlegel98} maps. As described in the introduction, it is well known that Schlegel maps are not trustable at regions closer to the galactic plane ($|b|<5^\circ$) due to contaminating sources and uncertainties in the dust temperatures. In order to quatify these effects, we compared our extinction values to a set of lines of sight at different latitudes. We retrieved $A_V$ values from the Schlegel maps for 1000 regions of $30' \times 30'$ size, randomly distributed across latitudes of $-8^\circ <b<-1^\circ$ for $0^\circ <l<2^\circ$. Although the relative difference between $A_{K_s}$ values from Schlegel maps and ours will not depend on the adopted extinction law, we converted their $A_V$ values into $A_{K_s}$ following \citet{cardelli89} law and then retrieved our own $A_{K_s}$ values from the BEAM calculator for the same target fields, again selecting the \citet{cardelli89} extinction law. Figure~\ref{schlegel} shows the comparison between our extinction results and those of \citet{schlegel98} as a function of latitude. The difference its already visible at $b<-6^\circ$ and becomes much larger for latitudes closer to the plane. Interestingly, we observe a negative difference between our $A_{K_s}$ values and those of \citet{schlegel98} for $b<-2^{\circ}$ while it becomes positive at $b>-2^{\circ}$. The source of this sign variation is not obvious given the uncertainties in the Schlegel maps at those latitudes.

\citet{marshall06} provide an extinction map for the complete Bulge, obtained by comparing the 2MASS photometry to the Besan\c{c}on model, including the model dependent extinction as a function of distance in each line of sight. Unfortunately, this technique only allows to calculate the maps with a lower resolution (15'), which is insufficient to resolve the small scale extinction variations in the inner Bulge regions.

Studies of \citet{schultheis99,schultheis09} provided very high resolution maps (2'), obtained using  DENIS and Spitzer data respectively for the inner $|b|<2^\circ$. Since in these regions the small scale extinction variations become particularly strong, these datasets are ideal to compare with our extinction maps and study possible systematics between the methods. The Spitzer map of \citet{schultheis09} covers specifically the $\sim 1^{\circ}$ of the very central regions of the Galaxy. This region is also included in the DENIS map and the extinction derived from DENIS and Spitzer observations agree very well.

In order to produce the DENIS extinction maps, \citet{schultheis99} used a method based on comparison of bulge RGB colors in 2' subfields with model isochrones from \citet{bertelli94}. They adopted a Solar metallicity population at a distance of 8 kpc with an age of 10 Gyr in the model isochrones. Same process was adopted to produce the Spitzer extinction maps from \citet{schultheis09}. This method has the benefit of using the bright Bulge stars, thus minimizing problems arising from photometric errors and (in-)completeness. However, the colors of RGB stars are much more strongly affected by population effects, in particular metallicity and to less extent age, when compared to RC stars in our method. In fact, the color of RGB stars is frequently used to derive the metallicity distribution function of stellar populations when spectroscopy of individual stars is not available. 

The influence of the population effects on the RC magnitudes and color has been investigated both theoretically \citep{girardi+salaris01,salaris+girardi02} and observationally \citep{alves00,grocholski+sarajedini02,pietrzynski+03,laney+12}. In both cases, RC stars appear as reliable standard distance indicator, considering the possible variations in the general Bulge population properties. If we consider a variation of the mean Bulge metallicity to go from Solar down to $\mathrm{[Fe/H]}=-0.4$, as obtained from the spectroscopic metallicity distributions of \citet{zoccali08}, we expect an error in the mean RC color of $\sim0.08$ mag, equivalent to a variation of $A_{K_s}=0.06$ mag assuming the \citet{cardelli89} extinction law and $A_{K_s}=0.04$ mag for \citet{nishiyama09} law. 

The extinction maps of \citet{schultheis99}, based on the magnitudes of RGB and AGB stars, were produced at latitudes $b<2^\circ$ where no or only very weak metallicity gradient seems to be present \citep{rich_origlia07}. Thus, extinctions obtained from isochrone comparison to red giant stars should be reliable for the regions studied by \citet{schultheis99} with errors dominated by the distance uncertainty of the RGB population towards each line of sight.  However, this method would have a considerable problem when used to construct a global extinction map, such us the one presented here.

\begin{figure}
\begin{center}
\includegraphics[width=0.40\textwidth]{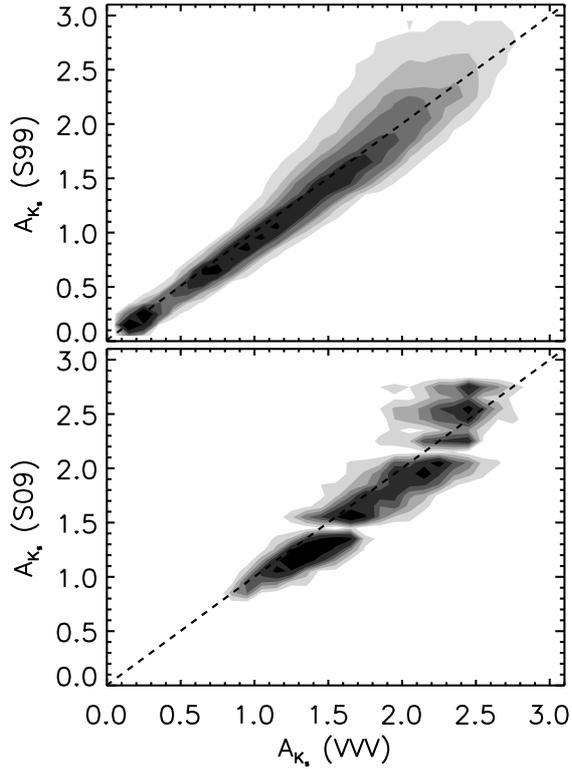}
\caption{Upper panel shows the comparison of the $A_{K_s}$ values obtained in this work and those of \citet{schultheis99} for the common inner Bulge region ($-2^\circ < b < +2^\circ$). Lower panel shows the comparison with the SPITZER extinction map of \citet{schultheis09} for the central $1^{\circ}$ of the Galaxy. Both plots are shown as density contours due to huge number of data points.}
\label{compplot}
\end{center}
\end{figure}

Figure~\ref{compmap} shows our extinction map and that of \citet{schultheis99} using DENIS photometry. $A_V$ values from \citet{schultheis99} were converted to $A_{K_s}$ following the extinction law of \citet{cardelli89} for comparison purposes. 
Their similarity is remarkable, with both maps tracing the same dust concentrations in the inner regions. We can also directly compare the extinction values retrieved from the BEAM calculator to those of \citet{schultheis99} and \citet{schultheis09}. This comparison is shown in Fig.~\ref{compplot}. $A_{K_s}$ values of both methods are in good agreement with a scatter of the order of 0.2 mag, which is expected from the different methods and photometric errors. The scatter becomes larger, up to 0.6 mag, in the regions of higher extinction ($A_{K_s}>2$), where for a significant number of lines of sight we find a lower extinction value than \citet{schultheis09}. 
A possible explanation for this difference is that, in these highly reddened regions, the extinction is enough to bring RC stars to magnitudes fainter than $K_s\sim17$ where the completness of VVV is of the order of $60\%$ \citep{saito12} and therefore our reddening values could be understimated. However, we emphasize that this difference is observed in the inner $1^\circ \times 1^\circ$ of the Bulge where extinction determinations are certainly more complicated. Moreover, the overall agreement is remarkable considering the different techniques, data sets and stellar populations (RC vs RGB) of our studies.

\section{Implications for Bulge studies}
\label{sect:implications}

Extinction corrections are a fundamental parameter for essentially all studies of stellar populations, as well as for distances and the determination of structural parameters. In this section we highlight the advantages provided by our new high resolution and large area extinction map for the studies of color magnitude diagrams, for determination of stellar parameters in spectroscopic abundance studies and for distance and structure studies of the Bulge.

\subsection{The Color-Magnitude diagram of the inner Bulge}

\begin{figure}
\begin{center}
\includegraphics[width=0.46\textwidth]{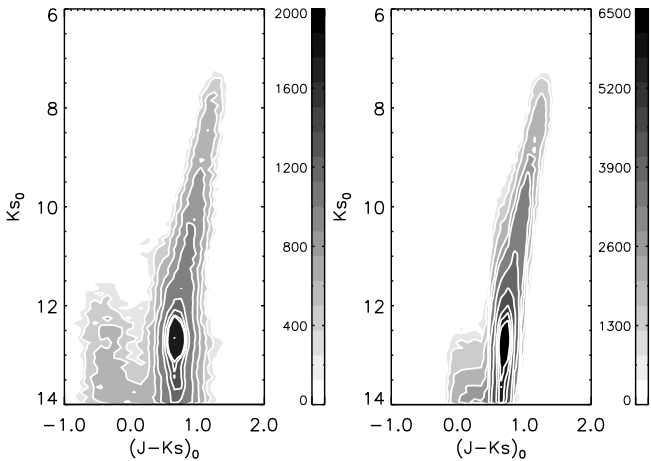}
\caption{Left panel shows HESS diagram of a 40' region in the inner Bulge $(l,b)=(-1.0,-1.0)$. The right panel shows the diagram for the same region as obtained from the Besancon model.}
\label{cmdcomp}
\end{center}
\end{figure}

The VVV color magnitude diagrams are de-reddened using our new extinction map of the Bulge. In Figure~\ref{cmdcomp} we compare the observed, extinction corrected, inner Bulge CMD with the predictions from the population synthesis model of the Milky Way from the Besan\c{c}on group including the thin disk, thick disk and bulge populations. This model is described in detail in \citet{robin12}. It uses the \citet{marshall06} extinction map and when compared with the observations such as 2MASS star counts it can be used to study the structure of the Milky Way. As already mentioned above, in the innermost regions of the Bulge, 2MASS is severely incomplete and affected by strong crowding and blending, thus limiting the predicting power of the model in the inner regions. With the superior spatial resolution of VVV survey (median seeing is better than 0.9" in $J$, $H$ and $K_s$ images) and its deeper photometry, it is now possible to make direct comparisons with the simulations and thus validate some of the assumptions adopted in the model.

Figure~\ref{cmdcomp} shows a Hess diagram of a $40' \times 40'$ region centered at Galactic coordinates 
$(l,b)=(1.0,-1.0)$ (VVV tile b320) in the left panel. In the right panel is the CMD obtained from the Besan\c{c}on model for the same region. A remarkable similarity is observed between our dereddened data and the model. The major difference is for the blue plume of stars at $(J-K_s)_0 \sim -0.2$ in the left diagram, which is much stronger, brighter, and bluer in the VVV CMD, than the blue sequence at $(J-K_s)_0 \sim 0.1$ in the model. This part of the CMD contains the disk stars, which are found all along the line of sight towards the Bulge, and therefore they have a range of reddening values, some having significantly lower reddening than derived for the Bulge red clump stars. Therefore after applying the extinction correction appropriate for the Bulge stars the colors of the disk stars cannot be correctly reproduced in our CMDs, but the colors and magnitudes of the major Bulge features, the RC and the RGB are in excellent agreement with the model.  This provides evidence that VVV dereddened color-magnitude diagrams can be used to investigate the stellar populations properties, such as age and metallicity distributions of stars even in the inner Bulge regions. Additionally, the good agreement between the model and the observations indicates that we could use the Besan\c{c}on model in comparison to observed VVV data, in order to obtain 3D extinction maps such as the one from \citet{marshall06}. The high spatial resolution and photometric depth of VVV CMDs offer the possibility to derive the 3D extinction map of the Bulge with a resolution of up to 6' for the inner regions (Chen et al. 2012, in preparation). Although this resolution, limited by the number of stars needed for the comparison, is lower than that of the map presented here, it will add the important 3D information, allowing more accurate characterization of the structure in the inner Bulge.

\subsection{Photometric effective temperatures}

Low and intermediate resolution spectroscopic studies, where determination of stellar parameters is based on photometric properties of the targets, depend on the adopted extinction corrections. In particular, the critical parameter is the effective temperature which is derived from the color of the stars.  Current studies of the Bulge use multi-object fibre spectrographs with large field of view in order to observe simultaneously large number of targets. The field of view can range from 25' radius for the FLAMES spectrograph on the VLT \citep{pasquini03} to   $2^\circ$ field of AAOmega multifibre spectrograph at the AAT \citep{sharp06}. Assuming that a single value of extinction is  adopted for a single pointing, using our 2' resolution extinction map we can estimate the differential extinction in a typical region of say 30' and derive the effect on the corresponding effective temperatures determined from colors of stars. 

\begin{figure}
\begin{center}
\includegraphics[scale=0.48]{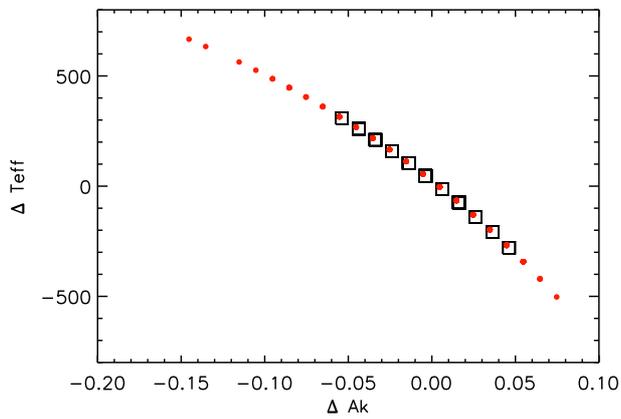}
\caption{The effect of differential reddening towards the Bulge for the determination of photometric effective temperatures. We plot the variations in the $A_{K_s}$ values for a set of 200 random positions in a 30' region towards $b=-4^\circ$ (black squares) and $b=-2^\circ$ (red circles), as obtained from the BEAM calculator, and compare them to variations in the photometric effective temperature.}
\label{teffplot}
\end{center}
\end{figure}

The relative effect of differential reddening will not depend significantly on the adopted calibration, and in our experiment we use the photometric calibration from Ramirez \& Melendez (2005) for giant stars.  A typical Bulge K giant star will have Solar metallicity with an effective temperature of around 4500 K. According to the calibration from \citet{ramirez05}, this star will have an intrinsic color of $(V-K_s)_0=2.625$.  We now consider a 30' region located at a latitude of $b=-4^\circ$ along the minor axis ($l=0^\circ$) and retrieve the extinction values using our BEAM calculator for a set of 200 randomly distributed positions within this region. These individual extinctions will differ from the mean extinction for the whole region depending on the amount of the differential extinction. If we add this difference to the intrinsic color of the star and use the new color to calculate the corresponding effective temperature from the photometric calibration, we obtain an estimation of the effect of differential reddening when adopting a single reddening value for the whole region. The difference between the original effective temperature and the one obtained after including the effect of differential extinction is shown in Figure~\ref{teffplot} as a function of the variation in $A_{K_s}$. The amount of extinction $A_{K_s}$ varies by 0.10 mag in the 30' region, which produces an error of up to 300 K in effective temperature. This effect becomes even more important in regions of higher extinction than Baade's Window as shown for the case at latitude $b=-2^\circ$ (red circles in Fig~\ref{teffplot}). Here a ~0.30 mag extinction variation is observed within the 30' region which implies errors larger than 600 K in the derived photometric effective temperature. 

Clearly all studies relying on photometric determination of the effective temperature need to take these extinction variations into account and correct for them. Otherwise the strong differential reddening in the inner Galaxy may result in large errors for the abundance measurements. While the dependence of metallicity error on temperature is not necessarily linear, and will depend on the spectral type of the target, for a reference we recall that already an error of 150 K in effective temperature led to a measurement error in [Fe/H] of 0.1 dex for K giant stars in the Bulge \citep{zoccali08}.

\subsection{Distances and the Bulge structure}

As shown in \citet{gonzalez11b} dereddened magnitudes of the VVV survey can be used to build the Bulge luminosity function towards different lines of sight and then measuring the mean magnitude of the RC to derive mean distance to different regions in the Bulge. This technique has been used in several studies to study the orientation angle of the Galactic bar \citep{stanek97, rattenbury07,cabrera07} and the X-shape morphology of the outer Bulge regions \citep{saito11}.

\begin{figure}
\begin{center}
\includegraphics[scale=0.55]{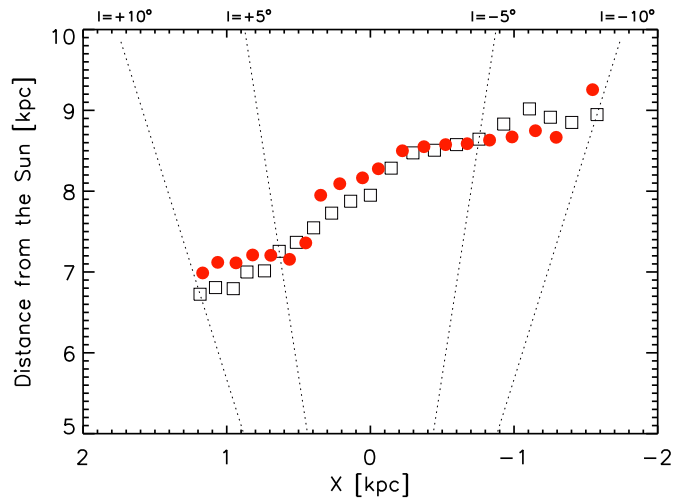}
\caption{Position of the Galactic bar with respect to the Sun as measured with the RC method from the the dereddened $K_s$ magnitudes using our extinction map (red filled circles) at a latitude of $b=-5^\circ$. Also shown, are the measurements from the Galactic model as presented in \citet{gerhard12} at the same latitude, corresponding to a bar with an angle of $25^{\circ}$ with respect to the Sun-Galactic center line of sight (black empty squares). Dotted lines mark the lines of sight for longitudes $l=\pm5^\circ$ and $l=\pm10^\circ$}
\label{rcplot}
\end{center}
\end{figure}

In \citet{gonzalez11c}  we studied the bar orientation in the inner regions ($b=\pm 1^\circ$) by measuring the mean RC magnitudes as a function of longitude. Using the homogeneous extinction map presented here, the study can now be extended to other regions of the Bulge. Ideally, these measurements should be compared to a galactic model such as the one of \citet{martinez06}. Interestingly, although this model has not been fine-tuned in any way to reproduce the Milky way properties, in \citet{gerhard12} the model was shown to nicely reproduce the observed structural properties of the Bulge, including the change in the bar inclination in the inner regions, as obtained by \citet{nishiyama05} and \citet{gonzalez11c}. This model also provides predictions for higher latitudes which can now be validated using our extinction-free magnitudes.  Figure~\ref{rcplot} shows position of the bar with respect to the Sun as measured from the mean dereddened $K_s$ VVV magntiudes and from the model of \citet{gerhard12} at higher Galactic latitude $b=-5^\circ$. The agreement between our observations and the model is remarkable, confirming that the Milky Way bar is inclined at 25 deg with respect to the Sun-Galactic center line of sight. The correction for differential extinction from our new map allows to extend this comparison of the model and observations to the complete Bulge, including the characterization of the stellar density profile, the observed X-shape, and other structural indicators.

\section{Conclusions}

We have used the VVV data to obtain a high-resolution extinction map of the Galactic bulge covering $\sim 315$ sq. deg area: $-10.3^\circ \leq b \leq +5.1^\circ$ and $-10.0^\circ \leq l \leq +10.4^\circ$. We used the color of RC stars to trace the effect of extinction in subfields of $2' \times 2'$ for  $-3.5^\circ \la b \la +5^\circ$,  $4' \times 4'$ for $-7^\circ \la b \la -3.5^\circ $ and $6' \times 6'$ for $-10^\circ \la b \la -7^\circ$, providing a first reddening map sensitive to the small scale variations and covering homogeneously the whole Bulge. 

An excellent agreement is found when comparing the common regions of our full VVV map with those of DENIS \citep{schultheis99} and Spitzer maps \citep{schultheis09} which cover the very inner regions of the Bulge ($|b|<2^\circ$). We also have compared our maps with those of \citet{schlegel98} at different latitudes and we confirm that Schlegel maps should not be used for latitudes $|b|<5^\circ$ towards the Milky Way bulge. 

The reddening map presented here becomes particularly important when the global picture of the Bulge is investigated. We presented some of the direct consequences of using a higher resolution map, when studying both the stellar populations and the structure of the Bulge. We show that a typical 30' region at $b=-4^{\circ}$, for which normally a single extinction value is assumed, has a variation of up to 0.10 magnitudes only due to differential reddening. This effect becomes even more important in the inner regions where differential reddening produces variations of more than 0.35 magnitudes in a similar field size. This has a strong impact in stellar population properties obtained from the color of stars, such as the photometric effective temperature. We show that errors of more than 500 K can be caused by differential extinction in a 30' region if single extinction value is adopted. The current map allows to avoid such errors. Also, we show that the observed CMD of an inner bulge region at $(l,b)=(1.0,-1.0)$ matches very well a CMD produced from the Besan\c{c}on stellar population model where effects of extinction are obviously not present.

We have shown how a good treatment of extinction and the dataset such as the one presented here, provides the ideal tool for the inner Galactic structure study based on the luminosity function properties. We used the VVV data to measure the mean RC magnitude of the Bulge towards different longitudes at $b=-5^\circ$, following the method described in \citep{gonzalez11}. We find a remarkable agreement between our measurements and those from the model of \citet{martinez06} for a bar at $25^{\circ}$ with respect to the Sun-Galactic center line of sight. \citet{gerhard12} have shown how the model reproduces our observations in the inner Bulge ($b=\pm1^\circ$) and we now show this is also true for larger latitudes. With our full extinction map it is possible to recover the global Bulge structural information in comparison to models such as the one of \citet{martinez06}.

Our extinction map is available to the community via a web based tool BEAM (Bulge Extinction And Metallicity) calculator (http://mill.astro.puc.cl/BEAM/calculator.php). The complete photometric metallicity map (Gonzalez et al. 2012 in prep.) will also be available from the same site. These maps are an important step towards the understanding of the general properties of the Galactic bulge and they are a particularly valuable tool for the forthcoming spectroscopic surveys of the inner Galaxy.     

%%%%%%%%%%%%%%%%%%%%%%%%%%%%%%%%%%%%%%%%%%%%%%%%%%%%%%%%%%%%%%%%%%%%%%%%%%%%%%%%
%% 

\begin{acknowledgements}
We thank Pascal Fouqu\'e for important suggestions to this work. We acknowledge the anonymous referee for helpfull comments on this article. We gratefully acknowledge use of data from the ESO Public Survey program ID 179.B-2002 taken with the VISTA telescope, data products from the Cambridge Astronomical Survey Unit. We acknowledge funding from the FONDAP Center for Astrophysics 15010003, the BASAL CATA Center for Astrophysics and Associated Technologies PFB-06, the MILENIO Milky Way Millennium Nucleus from the Ministry of Economycs ICM grant P07-021-F and Proyectos FONDECYT Regular 1110393 and 1090213. MZ acknowledges a Fellowship from the John Simon Guggenheim Memorial Foundation. MZ is also partially supported by Proyecto Anillo ACT-86. This publication makes use of data products from the Two Micron All Sky Survey, which is a joint project of the University of Massachusetts and Infrared Processing and Analysis Center/California Institute of Technology, funded by the National Aeronautics and Space Administration and the National Science Foundation. We warmly thank the ESO Paranal Observatory staff for performing the observations. 
%%%%%%%%%%%%%%%%%%%%%%%%%%%%%%%%%%%%% 
%%%%%%%%%%%%%%%%%%%%%%%%%%%%%%%%%%%%%
\end{acknowledgements}
\renewcommand*{\bibfont}{\small}
\bibliographystyle{aa}

\bibliography{mybiblio}

\begin{thebibliography}{57}
\expandafter\ifx\csname natexlab\endcsname\relax\def\natexlab#1{#1}\fi

\bibitem[{{Alves}(2000)}]{alves00}
{Alves}, D.~R. 2000, \apj, 539, 732

\bibitem[{{Babusiaux} {et~al.}(2010){Babusiaux}, {G{\'o}mez}, {Hill}, {Royer},
  {Zoccali}, {Arenou}, {Fux}, {Lecureur}, {Schultheis}, {Barbuy}, {Minniti}, \&
  {Ortolani}}]{babusiaux10}
{Babusiaux}, C., {G{\'o}mez}, A., {Hill}, V., {et~al.} 2010, \aap, 519, A77+

\bibitem[{{Bensby} {et~al.}(2011){Bensby}, {Ad{\'e}n}, {Mel{\'e}ndez}, {Gould},
  {Feltzing}, {Asplund}, {Johnson}, {Lucatello}, {Yee}, {Ram{\'{\i}}rez},
  {Cohen}, {Thompson}, {Bond}, {Gal-Yam}, {Han}, {Sumi}, {Suzuki}, {Wada},
  {Miyake}, {Furusawa}, {Ohmori}, {Saito}, {Tristram}, \& {Bennett}}]{bensby11}
{Bensby}, T., {Ad{\'e}n}, D., {Mel{\'e}ndez}, J., {et~al.} 2011, \aap, 533,
  A134

\bibitem[{{Bensby} {et~al.}(2010){Bensby}, {Feltzing}, {Johnson}, {Gould},
  {Ad{\'e}n}, {Asplund}, {Mel{\'e}ndez}, {Gal-Yam}, {Lucatello}, {Sana},
  {Sumi}, {Miyake}, {Suzuki}, {Han}, {Bond}, \& {Udalski}}]{bensby10}
{Bensby}, T., {Feltzing}, S., {Johnson}, J.~A., {et~al.} 2010, \aap, 512, A41+

\bibitem[{{Bertelli} {et~al.}(1994){Bertelli}, {Bressan}, {Chiosi}, {Fagotto},
  \& {Nasi}}]{bertelli94}
{Bertelli}, G., {Bressan}, A., {Chiosi}, C., {Fagotto}, F., \& {Nasi}, E. 1994,
  \aaps, 106, 275

\bibitem[{{Brown} {et~al.}(2010){Brown}, {Sahu}, {Anderson}, {Tumlinson},
  {Valenti}, {Smith}, {Jeffery}, {Renzini}, {Zoccali}, {Ferguson},
  {VandenBerg}, {Bond}, {Casertano}, {Valenti}, {Minniti}, {Livio}, \&
  {Panagia}}]{brown10}
{Brown}, T.~M., {Sahu}, K., {Anderson}, J., {et~al.} 2010, \apjl, 725, L19

\bibitem[{{Burstein} \& {Heiles}(1982)}]{bh82}
{Burstein}, D. \& {Heiles}, C. 1982, \aj, 87, 1165

\bibitem[{{Cabrera-Lavers} {et~al.}(2007){Cabrera-Lavers}, {Hammersley},
  {Gonz{\'a}lez-Fern{\'a}ndez}, {L{\'o}pez-Corredoira}, {Garz{\'o}n}, \&
  {Mahoney}}]{cabrera07}
{Cabrera-Lavers}, A., {Hammersley}, P.~L., {Gonz{\'a}lez-Fern{\'a}ndez}, C.,
  {et~al.} 2007, \aap, 465, 825

\bibitem[{{Cardelli} {et~al.}(1989){Cardelli}, {Clayton}, \&
  {Mathis}}]{cardelli89}
{Cardelli}, J.~A., {Clayton}, G.~C., \& {Mathis}, J.~S. 1989, \apj, 345, 245

\bibitem[{{Clarkson} {et~al.}(2008){Clarkson}, {Sahu}, {Anderson}, {Smith},
  {Brown}, {Rich}, {Casertano}, {Bond}, {Livio}, {Minniti}, {Panagia},
  {Renzini}, {Valenti}, \& {Zoccali}}]{clarkson08}
{Clarkson}, W., {Sahu}, K., {Anderson}, J., {et~al.} 2008, \apj, 684, 1110

\bibitem[{{Clarkson} {et~al.}(2011){Clarkson}, {Sahu}, {Anderson}, {Rich},
  {Smith}, {Brown}, {Bond}, {Livio}, {Minniti}, {Renzini}, \&
  {Zoccali}}]{clarkson11}
{Clarkson}, W.~I., {Sahu}, K.~C., {Anderson}, J., {et~al.} 2011, \apj, 735, 37

\bibitem[{{Elia Garcia Perez} {et~al.}(2012){Elia Garcia Perez}, {Allende
  Prieto}, {Bizyaev}, {Frinchaboy}, {Holtzman}, {Johnson}, {Majewski},
  {Nidever}, {Schiavon}, {Schultheis}, {Shetrone}, {Skrutskie}, {Wilson}, \&
  {Zasowski}}]{apogee12}
{Elia Garcia Perez}, A., {Allende Prieto}, C., {Bizyaev}, D., {et~al.} 2012, in
  American Astronomical Society Meeting Abstracts, Vol. 219, American
  Astronomical Society Meeting Abstracts, 410.04

\bibitem[{{Feltzing}(2011)}]{feltzing11}
{Feltzing}, S. 2011, in Stellar Clusters \& Associations: A RIA Workshop on
  Gaia, 311--320

\bibitem[{{Fulbright} {et~al.}(2007){Fulbright}, {McWilliam}, \&
  {Rich}}]{fulbright07}
{Fulbright}, J.~P., {McWilliam}, A., \& {Rich}, R.~M. 2007, \apj, 661, 1152

\bibitem[{{Gerhard} \& {Martinez-Valpuesta}(2012)}]{gerhard12}
{Gerhard}, O. \& {Martinez-Valpuesta}, I. 2012, \apjl, 744, L8

\bibitem[{{Girardi} \& {Salaris}(2001)}]{girardi+salaris01}
{Girardi}, L. \& {Salaris}, M. 2001, \mnras, 323, 109

\bibitem[{{Gonzalez} {et~al.}(2011{\natexlab{a}}){Gonzalez}, {Rejkuba},
  {Minniti}, {Zoccali}, {Valenti}, \& {Saito}}]{gonzalez11c}
{Gonzalez}, O.~A., {Rejkuba}, M., {Minniti}, D., {et~al.} 2011{\natexlab{a}},
  \aap, 534, L14

\bibitem[{{Gonzalez} {et~al.}(2011{\natexlab{b}}){Gonzalez}, {Rejkuba},
  {Zoccali}, {Hill}, {Battaglia}, {Babusiaux}, {Minniti}, {Barbuy},
  {Alves-Brito}, {Renzini}, {Gomez}, \& {Ortolani}}]{gonzalez11}
{Gonzalez}, O.~A., {Rejkuba}, M., {Zoccali}, M., {et~al.} 2011{\natexlab{b}},
  \aap, 530, A54+

\bibitem[{{Gonzalez} {et~al.}(2011{\natexlab{c}}){Gonzalez}, {Rejkuba},
  {Zoccali}, {Valenti}, \& {Minniti}}]{gonzalez11b}
{Gonzalez}, O.~A., {Rejkuba}, M., {Zoccali}, M., {Valenti}, E., \& {Minniti},
  D. 2011{\natexlab{c}}, \aap, 534, A3

\bibitem[{{Grocholski} \& {Sarajedini}(2002)}]{grocholski+sarajedini02}
{Grocholski}, A.~J. \& {Sarajedini}, A. 2002, \aj, 123, 1603

\bibitem[{{Hill} {et~al.}(2011){Hill}, {Lecureur}, {G{\'o}mez}, {Zoccali},
  {Schultheis}, {Babusiaux}, {Royer}, {Barbuy}, {Arenou}, {Minniti}, \&
  {Ortolani}}]{hill11}
{Hill}, V., {Lecureur}, A., {G{\'o}mez}, A., {et~al.} 2011, \aap, 534, A80

\bibitem[{{Indebetouw} {et~al.}(2005){Indebetouw}, {Mathis}, {Babler}, {Meade},
  {Watson}, {Whitney}, {Wolff}, {Wolfire}, {Cohen}, {Bania}, {Benjamin},
  {Clemens}, {Dickey}, {Jackson}, {Kobulnicky}, {Marston}, {Mercer},
  {Stauffer}, {Stolovy}, \& {Churchwell}}]{indebetouw05}
{Indebetouw}, R., {Mathis}, J.~S., {Babler}, B.~L., {et~al.} 2005, \apj, 619,
  931

\bibitem[{{Johnson} {et~al.}(2011){Johnson}, {Rich}, {Fulbright}, {Valenti}, \&
  {McWilliam}}]{johnson11}
{Johnson}, C.~I., {Rich}, R.~M., {Fulbright}, J.~P., {Valenti}, E., \&
  {McWilliam}, A. 2011, \apj, 732, 108

\bibitem[{{Kunder} {et~al.}(2008){Kunder}, {Popowski}, {Cook}, \&
  {Chaboyer}}]{kunder08}
{Kunder}, A., {Popowski}, P., {Cook}, K.~H., \& {Chaboyer}, B. 2008, \aj, 135,
  631

\bibitem[{{Laney} {et~al.}(2012){Laney}, {Joner}, \&
  {Pietrzy{\'n}ski}}]{laney+12}
{Laney}, C.~D., {Joner}, M.~D., \& {Pietrzy{\'n}ski}, G. 2012, \mnras, 419,
  1637

\bibitem[{{Lecureur} {et~al.}(2007){Lecureur}, {Hill}, {Zoccali}, {Barbuy},
  {G{\'o}mez}, {Minniti}, {Ortolani}, \& {Renzini}}]{lecureur07}
{Lecureur}, A., {Hill}, V., {Zoccali}, M., {et~al.} 2007, \aap, 465, 799

\bibitem[{{Marshall} {et~al.}(2006){Marshall}, {Robin}, {Reyl{\'e}},
  {Schultheis}, \& {Picaud}}]{marshall06}
{Marshall}, D.~J., {Robin}, A.~C., {Reyl{\'e}}, C., {Schultheis}, M., \&
  {Picaud}, S. 2006, \aap, 453, 635

\bibitem[{{Martinez-Valpuesta} {et~al.}(2006){Martinez-Valpuesta}, {Shlosman},
  \& {Heller}}]{martinez06}
{Martinez-Valpuesta}, I., {Shlosman}, I., \& {Heller}, C. 2006, \apj, 637, 214

\bibitem[{{McWilliam} \& {Rich}(1994)}]{mcwilliam94}
{McWilliam}, A. \& {Rich}, R.~M. 1994, \apjs, 91, 749

\bibitem[{{McWilliam} \& {Zoccali}(2010)}]{mcwilliam10}
{McWilliam}, A. \& {Zoccali}, M. 2010, \apj, 724, 1491

\bibitem[{{Minniti} {et~al.}(2010){Minniti}, {Lucas}, {Emerson}, {Saito},
  {Hempel}, {Pietrukowicz}, {Ahumada}, {Alonso}, {Alonso-Garcia}, {Arias},
  {Bandyopadhyay}, {Barb{\'a}}, {Barbuy}, {Bedin}, {Bica}, {Borissova},
  {Bronfman}, {Carraro}, {Catelan}, {Clari{\'a}}, {Cross}, {de Grijs},
  {D{\'e}k{\'a}ny}, {Drew}, {Fari{\~n}a}, {Feinstein}, {Fern{\'a}ndez
  Laj{\'u}s}, {Gamen}, {Geisler}, {Gieren}, {Goldman}, {Gonzalez}, {Gunthardt},
  {Gurovich}, {Hambly}, {Irwin}, {Ivanov}, {Jord{\'a}n}, {Kerins}, {Kinemuchi},
  {Kurtev}, {L{\'o}pez-Corredoira}, {Maccarone}, {Masetti}, {Merlo},
  {Messineo}, {Mirabel}, {Monaco}, {Morelli}, {Padilla}, {Palma}, {Parisi},
  {Pignata}, {Rejkuba}, {Roman-Lopes}, {Sale}, {Schreiber}, {Schr{\"o}der},
  {Smith}, {Sodr{\'e}}, {Soto}, {Tamura}, {Tappert}, {Thompson}, {Toledo},
  {Zoccali}, \& {Pietrzynski}}]{vvv10}
{Minniti}, D., {Lucas}, P.~W., {Emerson}, J.~P., {et~al.} 2010, \na, 15, 433

\bibitem[{{Nataf} {et~al.}(2010){Nataf}, {Udalski}, {Gould}, {Fouqu{\'e}}, \&
  {Stanek}}]{nataf10}
{Nataf}, D.~M., {Udalski}, A., {Gould}, A., {Fouqu{\'e}}, P., \& {Stanek},
  K.~Z. 2010, \apjl, 721, L28

\bibitem[{{Ness} \& {Freeman}(2012)}]{ness12}
{Ness}, M. \& {Freeman}, K. 2012, Assembling the Puzzle of the Milky Way, Le
  Grand-Bornand, France, Edited by C.~Reyl{\'e}; A.~Robin; M.~Schultheis; EPJ
  Web of Conferences, Volume 19, id.06003, 19, 6003

\bibitem[{{Nishiyama} {et~al.}(2005){Nishiyama}, {Nagata}, {Baba}, {Haba},
  {Kadowaki}, {Kato}, {Kurita}, {Nagashima}, {Nagayama}, {Murai}, {Nakajima},
  {Tamura}, {Nakaya}, {Sugitani}, {Naoi}, {Matsunaga}, {Tanab{\'e}},
  {Kusakabe}, \& {Sato}}]{nishiyama05}
{Nishiyama}, S., {Nagata}, T., {Baba}, D., {et~al.} 2005, \apjl, 621, L105

\bibitem[{{Nishiyama} {et~al.}(2006){Nishiyama}, {Nagata}, {Kusakabe},
  {Matsunaga}, {Naoi}, {Kato}, {Nagashima}, {Sugitani}, {Tamura}, {Tanab{\'e}},
  \& {Sato}}]{nishiyama06}
{Nishiyama}, S., {Nagata}, T., {Kusakabe}, N., {et~al.} 2006, \apj, 638, 839

\bibitem[{{Nishiyama} {et~al.}(2008){Nishiyama}, {Nagata}, {Tamura}, {Kandori},
  {Hatano}, {Sato}, \& {Sugitani}}]{nishiyama08}
{Nishiyama}, S., {Nagata}, T., {Tamura}, M., {et~al.} 2008, \apj, 680, 1174

\bibitem[{{Nishiyama} {et~al.}(2009){Nishiyama}, {Tamura}, {Hatano}, {Kato},
  {Tanab{\'e}}, {Sugitani}, \& {Nagata}}]{nishiyama09}
{Nishiyama}, S., {Tamura}, M., {Hatano}, H., {et~al.} 2009, \apj, 696, 1407

\bibitem[{{Ortolani} {et~al.}(1995){Ortolani}, {Renzini}, {Gilmozzi},
  {Marconi}, {Barbuy}, {Bica}, \& {Rich}}]{ortolani+95}
{Ortolani}, S., {Renzini}, A., {Gilmozzi}, R., {et~al.} 1995, \nat, 377, 701

\bibitem[{{Pasquini} {et~al.}(2003){Pasquini}, {Alonso}, {Avila}, {Barriga},
  {Biereichel}, {Buzzoni}, {Cavadore}, {Cumani}, {Dekker}, {Delabre}, {Kaufer},
  {Kotzlowski}, {Hill}, {Lizon}, {Nees}, {Santin}, {Schmutzer}, {Kesteren}, \&
  {Zoccali}}]{pasquini03}
{Pasquini}, L., {Alonso}, J., {Avila}, G., {et~al.} 2003, in Presented at the
  Society of Photo-Optical Instrumentation Engineers (SPIE) Conference, Vol.
  4841, Society of Photo-Optical Instrumentation Engineers (SPIE) Conference
  Series, ed. {M.~Iye \& A.~F.~M.~Moorwood}, 1682--1693

\bibitem[{{Pietrzy{\'n}ski} {et~al.}(2003){Pietrzy{\'n}ski}, {Gieren}, \&
  {Udalski}}]{pietrzynski+03}
{Pietrzy{\'n}ski}, G., {Gieren}, W., \& {Udalski}, A. 2003, \aj, 125, 2494

\bibitem[{{Ram{\'{\i}}rez} \& {Mel{\'e}ndez}(2005)}]{ramirez05}
{Ram{\'{\i}}rez}, I. \& {Mel{\'e}ndez}, J. 2005, \apj, 626, 446

\bibitem[{{Rattenbury} {et~al.}(2007){Rattenbury}, {Mao}, {Sumi}, \&
  {Smith}}]{rattenbury07}
{Rattenbury}, N.~J., {Mao}, S., {Sumi}, T., \& {Smith}, M.~C. 2007, \mnras,
  378, 1064

\bibitem[{{Rich} {et~al.}(2007){Rich}, {Origlia}, \&
  {Valenti}}]{rich_origlia07}
{Rich}, R.~M., {Origlia}, L., \& {Valenti}, E. 2007, \apjl, 665, L119

\bibitem[{{Robin} {et~al.}(2012){Robin}, {Marshall}, {Schultheis}, \&
  {Reyl{\'e}}}]{robin12}
{Robin}, A.~C., {Marshall}, D.~J., {Schultheis}, M., \& {Reyl{\'e}}, C. 2012,
  \aap, 538, A106

\bibitem[{{Saito} {et~al.}(2012){Saito}, {Hempel}, {Minniti}, {Lucas},
  {Rejkuba}, {Toledo}, {Gonzalez}, {Alonso-Garc{\'{\i}}a}, {Irwin},
  {Gonzalez-Solares}, {Hodgkin}, {Lewis}, {Cross}, {Ivanov}, {Kerins},
  {Emerson}, {Soto}, {Am{\^o}res}, {Gurovich}, {D{\'e}k{\'a}ny}, {Angeloni},
  {Beamin}, {Catelan}, {Padilla}, {Zoccali}, {Pietrukowicz}, {Moni Bidin},
  {Mauro}, {Geisler}, {Folkes}, {Sale}, {Borissova}, {Kurtev}, {Ahumada},
  {Alonso}, {Adamson}, {Arias}, {Bandyopadhyay}, {Barb{\'a}}, {Barbuy},
  {Baume}, {Bedin}, {Bellini}, {Benjamin}, {Bica}, {Bonatto}, {Bronfman},
  {Carraro}, {Chen{\`e}}, {Clari{\'a}}, {Clarke}, {Contreras}, {Corvill{\'o}n},
  {de Grijs}, {Dias}, {Drew}, {Fari{\~n}a}, {Feinstein},
  {Fern{\'a}ndez-Laj{\'u}s}, {Gamen}, {Gieren}, {Goldman},
  {Gonz{\'a}lez-Fern{\'a}ndez}, {Grand}, {Gunthardt}, {Hambly}, {Hanson},
  {He{\l}miniak}, {Hoare}, {Huckvale}, {Jord{\'a}n}, {Kinemuchi}, {Longmore},
  {L{\'o}pez-Corredoira}, {Maccarone}, {Majaess}, {Mart{\'{\i}}n}, {Masetti},
  {Mennickent}, {Mirabel}, {Monaco}, {Morelli}, {Motta}, {Palma}, {Parisi},
  {Parker}, {Pe{\~n}aloza}, {Pietrzy{\'n}ski}, {Pignata}, {Popescu}, {Read},
  {Rojas}, {Roman-Lopes}, {Ruiz}, {Saviane}, {Schreiber}, {Schr{\"o}der},
  {Sharma}, {Smith}, {Sodr{\'e}}, {Stead}, {Stephens}, {Tamura}, {Tappert},
  {Thompson}, {Valenti}, {Vanzi}, {Walton}, {Weidmann}, \&
  {Zijlstra}}]{saito12}
{Saito}, R.~K., {Hempel}, M., {Minniti}, D., {et~al.} 2012, \aap, 537, A107

\bibitem[{{Saito} {et~al.}(2011){Saito}, {Zoccali}, {McWilliam}, {Minniti},
  {Gonzalez}, \& {Hill}}]{saito11}
{Saito}, R.~K., {Zoccali}, M., {McWilliam}, A., {et~al.} 2011, \aj, 142, 76

\bibitem[{{Salaris} \& {Girardi}(2002)}]{salaris+girardi02}
{Salaris}, M. \& {Girardi}, L. 2002, \mnras, 337, 332

\bibitem[{{Schlegel} {et~al.}(1998){Schlegel}, {Finkbeiner}, \&
  {Davis}}]{schlegel98}
{Schlegel}, D.~J., {Finkbeiner}, D.~P., \& {Davis}, M. 1998, \apj, 500, 525

\bibitem[{{Schultheis} {et~al.}(1999){Schultheis}, {Ganesh}, {Simon}, {Omont},
  {Alard}, {Borsenberger}, {Copet}, {Epchtein}, {Fouqu{\'e}}, \&
  {Habing}}]{schultheis99}
{Schultheis}, M., {Ganesh}, S., {Simon}, G., {et~al.} 1999, \aap, 349, L69

\bibitem[{{Schultheis} {et~al.}(2009){Schultheis}, {Sellgren},
  {Ram{\'{\i}}rez}, {Stolovy}, {Ganesh}, {Glass}, \& {Girardi}}]{schultheis09}
{Schultheis}, M., {Sellgren}, K., {Ram{\'{\i}}rez}, S., {et~al.} 2009, \aap,
  495, 157

\bibitem[{{Sharp} {et~al.}(2006){Sharp}, {Saunders}, {Smith}, {Churilov},
  {Correll}, {Dawson}, {Farrel}, {Frost}, {Haynes}, {Heald}, {Lankshear},
  {Mayfield}, {Waller}, \& {Whittard}}]{sharp06}
{Sharp}, R., {Saunders}, W., {Smith}, G., {et~al.} 2006, in Society of
  Photo-Optical Instrumentation Engineers (SPIE) Conference Series, Vol. 6269,
  Society of Photo-Optical Instrumentation Engineers (SPIE) Conference Series

\bibitem[{{Stanek} {et~al.}(1994){Stanek}, {Mateo}, {Udalski}, {Szymanski},
  {Kaluzny}, \& {Kubiak}}]{stanek94}
{Stanek}, K.~Z., {Mateo}, M., {Udalski}, A., {et~al.} 1994, \apjl, 429, L73

\bibitem[{{Stanek} {et~al.}(1997){Stanek}, {Udalski}, {Szymanski}, {Kaluzny},
  {Kubiak}, {Mateo}, \& {Krzeminski}}]{stanek97}
{Stanek}, K.~Z., {Udalski}, A., {Szymanski}, M., {et~al.} 1997, \apj, 477, 163

\bibitem[{{Sumi}(2004)}]{sumi04}
{Sumi}, T. 2004, \mnras, 349, 193

\bibitem[{{Taylor}(2006)}]{taylor06}
{Taylor}, M.~B. 2006, in Astronomical Society of the Pacific Conference Series,
  Vol. 351, Astronomical Data Analysis Software and Systems XV, ed.
  {C.~Gabriel, C.~Arviset, D.~Ponz, \& S.~Enrique}, 666--+

\bibitem[{{Zoccali} {et~al.}(2008){Zoccali}, {Hill}, {Lecureur}, {Barbuy},
  {Renzini}, {Minniti}, {G{\'o}mez}, \& {Ortolani}}]{zoccali08}
{Zoccali}, M., {Hill}, V., {Lecureur}, A., {et~al.} 2008, \aap, 486, 177

\bibitem[{{Zoccali} {et~al.}(2003){Zoccali}, {Renzini}, {Ortolani}, {Greggio},
  {Saviane}, {Cassisi}, {Rejkuba}, {Barbuy}, {Rich}, \& {Bica}}]{zoccali03}
{Zoccali}, M., {Renzini}, A., {Ortolani}, S., {et~al.} 2003, \aap, 399, 931

\end{thebibliography}

\end{document}